\documentclass[11pt,preprint]{aastex}

\def\iso#1#2{\mbox{${}^{#2}{\rm #1}$}}
\def\he#1{\iso{He}{#1}}
\def\be#1{\iso{Be}{#1}}
\def\bor1#1{\iso{B}{1#1}}
\def\li#1{\iso{Li}{#1}}
\def\c1#1{\iso{C}{1#1}}
\def\n1#1{\iso{N}{1#1}}
\def\o1#1{\iso{O}{1#1}}

\def\gamflux{\rm photons \ cm^{-2} \ s^{-1} \ sr^{-1}}

\def\beq{\begin{equation}}
\def\eeq{\end{equation}}
\def\beqar{\begin{eqnarray}}
\def\eeqar{\end{eqnarray}}
\def\pfrac#1#2{\left( \frac{#1}{#2} \right)}
\def\pref#1{(\ref{#1})}

\def\pd{\partial}

\def\la{\mathrel{\mathpalette\fun <}}

\def\fun#1#2{\lower3.6pt\vbox{\baselineskip0pt\lineskip.9pt
  \ialign{$\mathsurround=0pt#1\hfil##\hfil$\crcr#2\crcr\sim\crcr}}}

\begin{document}

\title{Lithium-6 and Gamma Rays:  \\
Complementary Constraints on Cosmic-Ray History}

\author{Brian D. Fields\altaffilmark{1}
\altaffiltext{1}{also Department of Physics, University of Illinois}
}

\and

\author{Tijana Prodanovi\'{c}}

\affil{Center for Theoretical Astrophysics,
Department of Astronomy, University of Illinois,
Urbana, IL 61801}

\begin{abstract}

The rare isotope
\li6 is made only by cosmic rays, dominantly in
$\alpha\alpha \rightarrow \li6$
fusion reactions with ISM helium.
Consequently, this
nuclide provides a unique diagnostic of the history of cosmic rays 
in our Galaxy.  The same hadronic cosmic-ray interactions also produce
high-energy $\gamma$ rays
(mostly via $pp \rightarrow \pi^0 \rightarrow \gamma \gamma$).
Thus, hadronic $\gamma$-rays and \li6 are intimately linked.
Specifically, \li6 directly encodes the local cosmic-ray fluence
over cosmic time, while extragalactic hadronic $\gamma$ rays
encode an average cosmic-ray fluence over lines of sight out
to the horizon.
We examine this link and show how \li6 and $\gamma$-rays
can be used together to place important model-independent 
limits on the cosmic-ray
history of our Galaxy and the universe.
We first constrain $\gamma$-ray production from
ordinary Galactic cosmic rays, using the 
local \li6 abundance.
We find that the solar \li6 abundance
demands an accompanying extragalactic pionic $\gamma$-ray intensity
which exceeds that of the {\em entire} observed EGRB
by a factor of $2-6$.  Possible explanations for
this discrepancy are discussed.   
We then constrain Li production
using recent determinations of extragalactic $\gamma$-ray 
background (EGRB).
We note that cosmic rays created during cosmic
structure formation would lead to pre-Galactic Li production,
which would act as a ``contaminant'' to the
primordial \li7 content of metal-poor halo stars; 
the EGRB  can place an upper limit on this
contamination if
we attribute the entire EGRB pionic contribution 
to structure forming cosmic rays.
Unfortunately, the uncertainties in the determination of
the EGRB are so large that the present
$\gamma$-ray data cannot guarantee that
the pre-Galactic Li is small compared to primordial \li7;
thus, an improved determination of the EGRB will
shed important new light on this issue.
Our limits and their more model-dependent
extensions will improve significantly with
additional observations of \li6 in halo stars,
and with improved measurements of the EGRB spectrum
by GLAST.

\end{abstract}


\section{Introduction}

The origin and history of cosmic rays
has been a subject of
intensifying  interest.
For more than a decade, 
a large body of work has focused on the light
elements Li, Be, and B (LiBeB)
as signatures of cosmic-ray interactions
with the diffuse gas
\citep[for a recent review see][]{cva}.
LiBeB abundances in Galactic halo stars
have been used to probe the history of
cosmic rays in the (proto-)Galaxy.
More recently, a great deal of attention has
been focused on high-energy $\gamma$-rays
also produced in  interactions during cosmic-ray
propagation.
Here, we draw attention to the tight
connection between these observables, 
particularly between $\gamma$-rays and \li6.

The abundances of LiBeB nuclei 
encode the history of cosmic ray
exposure in local matter.
In the past 15 years or so, measurements of LiBeB in the Sun and
in Galactic disk have been joined by LiBeB observations
in halo stars; these offer particularly 
valuable information about cosmic-ray origins and 
interactions in Galactic and proto-Galactic matter.
In particular, different scenarios for cosmic ray origin
lead to different LiBeB trends, which have
been modeled and compared with observations
\citep[see, e.g., ][and references therein]{vc,fo99a,rslk}.
For the purposes of this paper, the details
of these models are less important than the following basic
distinction:
{\em all} LiBeB species are produced
as cosmic rays interact
with interstellar gas and
fragment--``spall''--heavy nuclei, e.g., $p + {\rm O} \rightarrow \be9$.
However, 
the fusion processes
$\alpha + \alpha \rightarrow \li{6,7}$
yield lithium isotopes exclusively, and indeed
dominate the cosmic-ray production of Li
\citep{sw,montmerle77c}. 
This makes cosmic-ray lithium production particularly 
``clean'' since its evolution depends uniquely on its
exposure to cosmic rays, and unlike Be and B, does {\em not} depend
on the ambient heavy element abundances.  
 
Cosmic-ray interactions provide the {\em only} known source
for the nucleosynthesis of \li6, \be9, and \bor10,
making these species ideal observables of cosmic ray 
activity.\footnote{
In fact, a pre-Galactic component of \li6
can be produced in some scenarios in which 
dark matter decays via hadronic \citep{dimopoulos}
or electromagnetic \citep{karsten,kkm,cefo} channels.
Such scenarios are constrained via their effects
on the other light elements, but some level
of \li6 production is hard to rule out completely.
}  
The story is more complex for
\bor11, which can also be produced in core-collapse supernovae
by the ``neutrino process'' \citep[e.g.,][]{whhh,ytks}.
Finally, \li7 has the most diverse lineage.
In the early Galaxy, and hence in halo stars, \li7 is dominated by 
the contribution from primordial nucleosynthesis
\citep[e.g.,][and references therein]{cfo},
with a small contribution from cosmic-ray fusion
as well as the neutrino process \citep{rbofn}.
At late times, and hence in disk stars including the
Sun, \li7 has important and probably dominant contributions
from longer-lived, low-mass stars 
\citep[though the specific site remains controversial:][]{rmvd,claudia}.
In this paper we will
build on the work of 
\citet{si} to point out the possible importance of
another, {\em pre-Galactic}, source of cosmic-ray \li7 and \li6,
which could confound attempts to identify the pre-Galactic
Li abundance with the primordial component.
We cannot rule out (or in!) this possible source,
but we will constrain it using
observations of $\gamma$-rays.

Cosmic-ray interactions with interstellar gas
produce not only LiBeB, but also inevitably produce
$\gamma$-rays.  Cosmic rays in the Galactic disk today
lead to pronounced emission seen in the 
Galactic plane \citep{hunter}.
Cosmic ray populations in (and between!) external galaxies
would contribute to
a diffuse extragalactic $\gamma$-ray background
(hereafter the EGRB).
The existence of an EGRB 
was already claimed by some of the first $\gamma$-ray
observations \citep{fkh}.
The most recent high-energy (i.e., roughly 
in the 30 MeV -- 30 GeV range) $\gamma$-ray observations are those of
the EGRET experiment on the Compton Gamma-Ray Observatory,
and the EGRET team also found
evidence for a EGRB  \citep{sreekumar}.
The intensity, energy spectrum, and even the existence of
an EGRB are not trivial to measure, as this information
only arises as the residual 
after subtracting the dominant
Galactic foreground from the observed
$\gamma$-ray sky.
The procedure for foreground subtraction is thus crucial,
and different procedures starting with the same EGRET data
have arrived at an EGRB with a lower intensity and
different spectrum \citep{smr}, or have even failed
to find evidence for an EGRB at all \citep{kwl}.
Despite these uncertainties, we will see that the
EGRB (or limits to it) and Li abundances are mutually
very constraining.

Whether or not an EGRB has yet been detected,
at some level it certainly should exist.
EGRET detections of individual active galactic nuclei (blazars)
as well as the Milky Way and the LMC
together guarantee that unresolved
blazars \citep[e.g.,][]{ss,mc},
and to a lesser extent normal galaxies \citep{pf},
will generate a signal at or near
the levels claimed for the EGRB.
Many other EGRB sources  have been
proposed, but one of the promising
has been a subject of intense interest recently: namely,
$\gamma$-rays originating from a cosmological component of
cosmic rays. 
This as-yet putative cosmic-ray population
would originate in shocks 
\citep{mrkjco,kwlsh,rkhj}
associated with baryonic infall and merger events
during  
the growth of large-scale cosmic structures.
Diffusive shock acceleration
\citep[e.g.,][]{blasi,kjg,je}
would then generate a population of
relativistic ions and electrons.
Gamma-ray emission would then follow from
inverse Compton scattering of electrons off
of the ambient photon backgrounds
and from $\pi^0$ production due to hadronic collisions
\citep{lw}.
The most recent semi-analytical and numerical calculations 
\citep{gb,miniati}
suggest that this ``structure forming'' component to
the EGRB is likely below the blazar contribution,
but the observational and theoretical uncertainties
here remain large; upcoming $\gamma$-ray observations 
by GLAST 
\citep{glast} 
will shed welcome new light on this problem.

The link between the nucleosynthesis and
$\gamma$-ray signatures of cosmic-ray history
has been pointed out by others in multiple contexts.
We note in particular the prescient work of
\citet{montmerle77a,montmerle77b,montmerle77c},
who in a series of papers considered
the implications of a hypothetical population of ``cosmological cosmic
rays'' in addition to the usual Galactic cosmic rays.
Montmerle's analysis is impressive in its foresight
and its breadth.   
\citet{montmerle77a} develops the formalism
for a homogeneous population of cosmological
cosmic rays (assumed to be created instantaneously at some
redshift), and describes their propagation in an
expanding universe, as well as their light element
and $\gamma$-ray production.
He identifies the tight connection between
\li6 and extragalactic $\gamma$-rays, and 
exploits this connection to use the available 
EGRB data to constrain Li production for a variety
of different assumptions.  
A particularly pertinent case involves
an EGRB near the levels discussed today
(``normalization 2'' in Montmerle's parlance),
coupled with a cosmic baryon density 
close to modern values \citep[e.g.,][]{wmap,cfo}.
Under these conditions, \citet{montmerle77b}
finds that cosmological cosmic ray activity
sufficient to explain the EGRB
leads to a present \li6 abundance that is
about an order of magnitude smaller than the
solar abundance.  This result foreshadows
an important conclusion we will find:
if the solar \li6 abundance is produced by
Galactic cosmic rays, then 
the associated pionic $\gamma$-ray production
exceeds the {\em entire} EGRB by at least
a factor of 2.

More recently, studies of structure formation
cosmic rays have focused primarily on their 
$\gamma$-ray signatures.  However, recently
\citet{si} also proposed using \li6 as a diagnostic
of shock activity in the Local Group.
These authors note that the resulting \li6 abundances in halo
stars could be used to probe the shocks and 
resulting cosmic rays in proto-Galactic matter.
We also will draw on this idea, 
with an emphasis on the fact that pre-Galactic
Li production would be (by itself) observationally 
indistinguishable from the primordial \li7
production from big bang nucleosynthesis.
Thus we attempt to use EGRB data to constrain
this possibility, but find that current data
is unable to rule out a significant  
contribution to halo star Li abundances
from structure formation cosmic rays.

Our work thus follows these pioneering efforts,
further emphasizing and formally exploring the
intimate connection between cosmic-ray nucleosynthesis
and high-energy $\gamma$-ray astrophysics.
In \S \ref{sect:gamma-li6},
we formally show and discuss the generality and tightness of the 
$\li6$-$\gamma$ connection, and 
in \S \ref{sect:obs} the relevant observations are reviewed.
In \S \ref{sect:gcr} we use the theory of LiBeB
production by Galactic cosmic rays to 
deduce the minimal contribution the EGRB.
In \S \ref{sect:sfcr} we exploit the link to use the
observed EGRB to limit the SFCR contribution to
pre-Galactic lithium.
Discussion and conclusions appear in \S \ref{sect:discuss}.

\section{The Gamma-Ray -- Lithium Connection:  Formalism}
\label{sect:gamma-li6}

Before doing a detailed calculation let us first establish a simple, 
back of the envelope, connection between gamma-rays and lithium. 
We know that low energy ($\sim 10-100$ MeV/nucleon) 
hadronic cosmic rays produce 
lithium through $ \alpha \alpha \rightarrow 
\li{6,7} + \cdots$.  But higher-energy 
($> 280$ MeV/nucleon) cosmic rays also produce $\gamma$-rays 
via neutral pion decay: $pp \rightarrow \pi^0 \rightarrow \gamma \gamma$.
Because they share a common origin in hadronic cosmic ray interactions,
we can directly relate cosmic ray lithium production to ``pionic'' 
gamma-rays.   The cosmic-ray production rate of $\li6$ per unit volume is
$q({\li6}) = \sigma_{\alpha \alpha \rightarrow {}^6{\rm Li}} 
\Phi_{\alpha} n_{\alpha}$,
where $\Phi_\alpha$  is the net cosmic ray He flux,
$n_\alpha$ is the interstellar He abundance, and 
$\sigma_{\alpha \alpha \rightarrow {}^6{\rm Li}}$ is the
cross section for \li6 production, 
appropriately averaged over
the cosmic-ray energy spectrum
(detailed definitions and normalization conventions appear
in Appendix \ref{sect:norm}).
Thus, the $\li6$ mole fraction $Y_6 = n_6/n_{\rm b}$ is just 
$Y({\li6}) \sim \int  \frac{dt}{n_b} q_{\li6}
  \sim y_{\alpha,\rm cr} Y_{\alpha,\rm ism} 
    \sigma_{\alpha \alpha \rightarrow {}^6{\rm Li}} \Phi_p t_0$.
where 
$y_{\alpha,\rm cr} = \Phi_\alpha/\Phi_p \approx {\rm (He/H)_{ism}}$. 

On the other hand, the
cosmic-ray production rate of 
pionic $\gamma$-rays is just the pion production rate
times a factor of 2, that is, 
$q_{\gamma} = 
  2 \sigma_{p p \rightarrow \pi^0} \Phi_{p,\rm cr} n_{p,\rm ism}$. 
Integrated over a line of sight towards the cosmic particle horizon, 
this gives a EGRB intensity
$I_{\gamma} \sim c \int dt q_{\gamma}/4\pi 
  \sim 2 \sigma_{p p \rightarrow \pi^0} c \Phi_p t_0$. 
Thus we see that both the \li6 abundance
and the $\gamma$-ray intensity have a common factor of the
(time-integrated) cosmic-ray flux,   and so
we can eliminate this factor and express each observable
in terms of the other:
\beq
Y_{\li6} 
\sim y_{\alpha,{\rm cr}} Y_{\alpha,{\rm gas}} \frac{2\pi}{n_b c}
\frac{\sigma_{\li6}^{\alpha \alpha}}
{\sigma_{\pi^0}^{p p}} I_{\gamma}
\label{eq:connection}
\eeq
From eq.~(\ref{eq:connection}) we see that the connection between cosmic-ray 
lithium production and pionic gamma-ray flux is straight forward.

This rough argument shows the intimacy of the connection
between \li6 and pionic $\gamma$-rays.  However,
this simplistic treatment does not account for 
the expansion of the universe, nor for time-variations
in the cosmic-cosmic ray flux, nor for the inhomogeneous
distribution of sources within the universe.
We now include these effects in a more rigorous treatment.

For Li production at location $\vec{x}$, the production rate
per unit (physical) volume is 
\beq
q_{\rm Li}(\vec{x}) 
 =  \sigma_{\alpha \alpha} \Phi_\alpha^{\rm cr}(\vec{x})  n_{\rm \alpha,gas}
(\vec{x}) 
 = y_{\alpha,\rm cr} Y_{\alpha}^{\rm ism}
   \sigma_{\alpha \alpha} \Phi_p^{\rm cr}(\vec{x}) n_{\rm b,gas}(\vec{x}) 
 \equiv \mu(\vec{x}) \Gamma_{\rm Li}(\vec{x})  n_{\rm b}(\vec{x})
\eeq
Here, $y_{\alpha,\rm cr}=(\alpha/p)_{\rm cr}$ is the cosmic-ray
He/H ratio, and is assumed to be constant in space and time.\footnote{
That is, we ignore the small non-primordial \he4 production by stars,
and we neglect any effects of H and He segregation.
Both of these should be quite reasonable approximations.
}
The target density of (interstellar or intergalactic) helium
is $n_{\rm \alpha,gas}$, which we write in terms of its
ratio $Y_\alpha^{\rm ism} = n_{\rm He}/n_{\rm b}$ to the baryon
density. We take $Y_\alpha^{\rm ism} \approx 0.06$ to be constant
in space and time, but we do not assume this
for the baryon density $n_{\rm b}(\vec{x})$.
The baryonic gas fraction
\beq
\mu = n_{\rm b,gas}/n_{\rm b}
\eeq
accounts for the fact that not all baryons need
to be in a diffuse form. 
Finally, we will find it convenient to write
$q_{\rm Li}(\vec{x})$  in terms of the 
local baryon density and the local Li production
rate $\Gamma_{\rm Li}(\vec{x})$ per baryon.

With these expressions, we have
\begin{equation}
\frac{d}{dt} Y_{\rm Li}(\vec{x}) = \mu(\vec{x}) \Gamma_{\rm Li}(\vec{x})
\end{equation}
which we can solve to get
\beqar
Y_{\rm Li}(\vec{x},t) 
  & = & \int_0^t dt^\prime \ \mu(\vec{x},t^\prime) \ \Gamma_{\rm Li}(\vec{x},t^\prime) \\
\label{eq:li}
  & = &  y_{\alpha,\rm cr} Y_{\alpha}^{\rm ism} \sigma_{\alpha \alpha} 
     \int_0^t dt^\prime \  \mu(\vec{x},t^\prime) \ \Phi_p^{\rm cr}(\vec{x},t^\prime) \\
  & = &  y_{\alpha,\rm cr} Y_{\alpha}^{\rm ism} \sigma_{\alpha \alpha} 
F_p({\vec x},t)
\eeqar
where $F_p(\vec{x},t) = \int_0^t dt^\prime \  \mu(\vec{x},t^\prime) \ 
\Phi_p^{\rm cr}(\vec{x},t^\prime)$ is
the local proton fluence (time-integrated flux), weighted
by the gas fraction.
Thus we see that Li (and particularly \li6) serves
as a ``cosmic-ray dosimeter'' which measures
the net local cosmic ray exposure.

We now turn to $\gamma$ rays from 
hadronic sources, most of which come
from neutral pion production and decay:
$pp \rightarrow \pi^0 \rightarrow \gamma \gamma$.
The extragalactic background due to these
process is expected to be isotropic
(at least to a good approximation).
In this case, 
the total $\gamma$-ray intensity 
$I_\gamma = dN_\gamma/dA \, dt \, d\Omega$,
integrated over all energies, is given by an integral
\beq
\label{eq:intensity}
I_\gamma(t) 
  = \frac{c}{4\pi} \int_0^t dt^\prime \ q_\gamma^{\rm com}(t^\prime)
\eeq
of the sources over the line of sight to the horizon.
We are interested in particular in the case of hadronic
sources, so that
$q_{\rm com} = a^3 q$ is the total (energy-integrated)
comoving rate of hadronic $\gamma$-ray production per unit volume;
here $a$ is the usual cosmic scale factor, which
we normalize to a present value of $a_0 = a(t_0) = 1$. 
A formal derivation of eq.\ \pref{eq:intensity}
appears in Appendix \ref{sect:radtransf}, but one can arrive at this 
result from elementary considerations.
Namely, note that the comoving number density of
photons produced at any point is just 
$n_{\gamma,\rm com} = \int_0^t   q_\gamma^{\rm com} \, dt^\prime$.
We neglect photon absorption and scattering processes,
and thus particle number conservation along with
homogeneity and isotropy together demand
that the comoving number density of ambient photons
at any point is the same as the comoving number density of
photons produced there.
Furthermore, the total (energy-integrated)
photon intensity is also isotropic and thus by definition is
$I_\gamma = n_{\gamma,\rm com} c/4\pi$,
which is precisely what we find in eq.\  \pref{eq:intensity}.

The comoving rate of pionic $\gamma$-ray production per unit volume 
at point $\vec{s}$ is
\beq
\label{eq:qgamma}
q_{\gamma}^{\rm com}(\vec{s},t) 
 = \sigma_{\gamma} \Phi_p(\vec{s},t) 
   n_{\rm H,gas}^{\rm com}(\vec{s},t)
 = \mu(\vec{s},t) \sigma_{\gamma} \Phi_p(\vec{s},t) 
   n_{\rm H}^{\rm com}(\vec{s},t)
\eeq
where $n_{\rm H}$ is the (comoving) hydrogen density,
and 
$\Phi_p = 4\pi \int I_p(\epsilon_{\rm rest}) d\epsilon_{\rm rest}$
is the total (integrated over rest-frame energy 
$\epsilon_{\rm rest}$) omnidirectional cosmic ray proton flux.
The flux-averaged pionic $\gamma$-ray production cross section 
is
\beq
\sigma_{\gamma} \equiv 2 \xi_\alpha \zeta_\pi \sigma_{\pi^0} 
 = 2 \xi_\alpha
  \frac{\int d\epsilon_{\rm rest}  \,
    I_p(\epsilon_{\rm rest}) \, \zeta_\pi \sigma_{\pi^0}(\epsilon_{\rm rest})}
       {\int d\epsilon_{\rm rest} \, I_p(\epsilon_{\rm rest})}
\eeq
where the factor of 2 counts the number of photons per pion decay,
$\sigma_{\pi^0}$ is the cross section for pion production
and $\zeta_\pi$ is the pion multiplicity, 
and the factor $\xi_\alpha  = 1.45$ accounts for 
$p\alpha$ and $\alpha\alpha$ reactions \citep{dermer}.

Then we have
\beq
I_{\gamma}(t)  = \frac{ n_{\rm b,0} c}{4\pi} 
  Y_{\rm H} \sigma_{\gamma} 
\int_0^t dt \ \mu(\vec{s}) \ \frac{n_{\rm b}^{\rm com}(\vec{s})}{n_{\rm b,0}} \ \Phi_p(\vec{s},t) 
 = \frac{ n_{\rm b,0}c}{4\pi} \sigma_{\gamma} Y_H^{\rm ism}F_p(t)
\eeq
where 
\beq	
F_p(t) 
 = \int_0^t dt \ \mu(\vec{s}) \ \frac{n_{\rm b}^{\rm com}(\vec{s})}{n_{\rm b,0}} \ \Phi_p(\vec{s}) 
\eeq
is a mean value of the cosmic-ray fluence along the line of sight,
where the average is
weighted by the gas fraction and the ratio 
$n_{\rm b}^{\rm com}(\vec{s})/n_{\rm b,0}$
of the local baryon density along the photon path.
Note that the $\gamma$-ray sources are sensitive to the overlap
of the cosmic-ray flux with the diffuse hydrogen gas density,
and thus need not 
be homogeneous.  Even so, we still assume the ERGB intensity to
be isotropic, which corresponds to the assumption that the
line-of-sight integral over the sources averages out
any fluctuations.

One further technical note:
$I_{\gamma} \equiv I_{\gamma}(>0) 
  = \int_0^\infty d\epsilon_\gamma I_{\gamma}(\epsilon_\gamma)$ 
represents the total pionic 
$\gamma$-ray flux, integrated over photon energies.
While this quantity is well-defined theoretically,
real observations have some energy cutoff,
and thus report
$I_{\gamma}(> \epsilon_0) 
  = \int_{\epsilon_0}^\infty d\epsilon_\gamma I_{\gamma}(\epsilon_\gamma)$,
typically with $\epsilon_0 = 100$ MeV.
But the 
spectrum of pionic $\gamma$-rays will be shifted towards lower energies if 
they originate from a nonzero
redshift. Thus it is clear that $\gamma$-ray intensity $I_{\gamma}$, 
integrated above some energy $\epsilon_0 \ne 0$,
will be redshift-dependent. A way to eliminate this $z$-dependence is to 
include {\it all} pionic
$\gamma$-rays, that is to take $I_{\gamma} \equiv I_{\gamma}(>0 \rm \ GeV)$,
i.e., to take $\epsilon_0 = 0$. 
As discussed in more detail in Appendix \ref{sect:radtransf},
the $\li6$-$\gamma$ proportionality is only
exact for $I_{\gamma}(> 0)$, as
this quantity removes photon redshifting effects
which spoil the proportionality for $\epsilon_0 \ne 0$.
Thus we will have to use information on
the pionic spectrum to translate between
$I_{\gamma}(> \epsilon_0)$
and $I_{\gamma}(> 0)$; these issues are discussed further
in \S \ref{sect:pionic}.

Thus we see that
the lithium abundance and the
pionic $\gamma$-ray intensity (spectrum integrated from 0 energy) arise from
very similar integrals, which we can express via
the ratio
\beq
\label{eq:gamma2li}
\frac{I_\gamma(t)}{Y_i(\vec{x},t)}
 = \frac{ n_{\rm b} c}{4\pi y_{\alpha ,\rm  cr} y_{\alpha ,\rm ism}} 	\
   \frac{ \sigma_{\gamma}}{\sigma_{\alpha \alpha}^i} \
   \frac{F_p(t)}{F_p(\vec{x},t)}   
\eeq
where $i$ denotes \li6 or \li7.
Note that this ``$\gamma$-to-lithium'' ratio
has its only significant space
and time dependence via
the ratio $F_p(t)/F_p(\vec{x},t)$ of the line-of-sight baryon-averaged fluence
to the local fluence.\footnote{
In fact, the ratio also 
depends on the shape of the cosmic-ray spectrum (assumed universal),
which determines the ratio of cross sections.
We will take this into account below when we
consider different cosmic-ray populations.
}

The relationship expressed in eq.\ \pref{eq:gamma2li}
is the main result of this paper, and we will
bring this tool to bear on Li and $\gamma$-ray observations,
using each to constrain the other.
To do this, it will be convenient to write
eq.\ \pref{eq:gamma2li} in the form
\beq
\label{eq:main}
I_\gamma(t)
 = I_{0,i} \frac{Y_i(\vec{x},t)} {Y_{i,\odot}}
  \frac{F_p(t)}{F_p(\vec{x},t)}   
\eeq
where the scaling factor
\beq
I_{0,i} = \frac{ n_{\rm b} c}{4\pi y_{\alpha,\rm cr} y_{\alpha ,\rm ism}} \
   \frac{ \sigma_{\gamma}}{\sigma_{\alpha \alpha}^i} \
   Y_{i,\odot}
\eeq
is independent of time and space, and only
depends, via the ratio of cross sections,
on the shape of the cosmic-ray population considered.
Table \ref{table:scaling}  presents the values of $I_{0,i}$ for
the different spectra that will be considered in the following sections.
 Values of the scaling factor
were obtained by using photon multiplicity $\xi_\gamma =2$,
$\zeta_{\alpha}=1.45$,  baryon number density $n_{\rm b}=2.52 \times 10^{-7} 
\ \rm cm^{-3}$, CR and ISM helium abundances 
$y_{\rm \alpha}^{\rm cr}=y_{\rm \alpha}^{\rm ism}=0.1$ and solar abundances
 $y_{\li6_\odot}=1.53 \times 10^{-10}$ and 
$y_{\li7_\odot}=1.89 \times 10^{-9}$ \citep{ag}. 
For the $\pi^0$ and lithium production
cross-sections, we used the fits taken from  \citet{dermer} and \citet{Mercer},
and from that obtained
the ratios of flux-averaged cross-sections 
for different spectra, and
these are also presented in Table \ref{table:scaling}.  

\begin{table}[h]
\epsscale{0.5}
\begin{center}
\caption {Lithium and $\gamma$-ray Scalings and Production Ratios
  \label{table:scaling}}
\begin{tabular}{|c|ccccc|}
\tableline\tableline
Cosmic-Ray & $\rm I_{0,6} $ & $\rm I_{0.7}$ & & & \\
Population & \multicolumn{2}{c}{$[\rm cm^{-2} s^{-1} sr^{-1}]$} &
  $\sigma_{\li6}^{\alpha \alpha}/\sigma_{\pi}^{pp}$  & 
  $\sigma_{\li7}^{\alpha \alpha}/\sigma_{\pi}^{pp}$  & 
  \li7/\li6 \\
\tableline
GCR & $9.06 \times 10^{-5} $ & $ 8.36 \times 10^{-4} $ & 0.21 & 0.28 & 1.3 \\
SFCR & $ 1.86 \times 10^{-5} $ & $ 1.15 \times 10^{-4} $ & 1.02 & 2.03 & 2.0 \\
\tableline\tableline
\end{tabular}
\end{center}
\end{table}

Table \ref{table:scaling} shows that the different
cosmic-ray spectra lead to
very different Li-to-$\gamma$ ratios.
For example, the \li6-to-$\gamma$ ratio
$\sigma_{\li6}^{\alpha \alpha}/\sigma_{\pi}^{pp}$
is almost a factor of 5 higher
in the SFCR case than in the GCR case.
The reason for this
stems from the different threshold behaviors
and energy dependences of the Li and $\pi^0$
production cross sections.
Li production via $\alpha\alpha$ fusion
has a threshold around 10 MeV/nucleon,
above which the cross section rapidly rises through
some resonant peaks.  Then beyond
$\sim 15$ MeV/nucleon, the cross section for \li6
{\em drops} exponentially as $e^{-E/{16 \ \rm MeV/nucleon}}
$\citep{Mercer},
rapidly suppressing the importance of any
projectiles with $E \gg 16$ MeV/nucleon.
Thus, as has been widely discussed, 
Li production is a low-energy phenomenon
for which the important projectile energy range is
roughly $10 - 70$ MeV/nucleon.

On the other hand, 
$pp \rightarrow \pi^0$ production has a higher threshold
of 280 MeV, and the effective cross section $\zeta_\pi \sigma_{pp}^\pi$
{\em rises} with energy up to and beyond 1 GeV.
Neutral pion production is thus a significantly higher-energy
phenomenon.  

These different cross section behaviors
are sensitive to the differences in the two 
cosmic-ray spectra we adopt.
On the one hand, we adopt a GCR spectrum 
that is a power law in {\em total} energy:
$\phi_p(E) \propto (m_p + E)^{-2.75}$,
a good approximation to the 
locally observed (i.e., {\em propagated})
spectrum.
This spectrum is roughly constant for
$E < m_p$. Thus, there is no reduction in
cosmic-ray flux between the Li and $\pi^0$
thresholds.  Furthermore the flux only begins to drop
far above the $\pi^0$ threshold at 280 MeV, so that
there is significant pion production over
a large range of energies, in contrast to the
intrinsically narrow energy window for 
Li production.  As a result of the effects,
$\sigma_{\li6}^{\alpha \alpha}/\sigma_{\pi}^{pp} \ll 1$
for the GCR case.

In contrast, the SFCR flux
is taken to be the standard result for
diffusive acceleration due to a strong
shock: namely, a  power law in momentum
$\phi(E) \propto p(E)^{-2}$.
This goes to $\phi \propto E^{-1}$ at
$E \la m_p$, and $\phi \propto E^{-2}$ at
higher energies.  This spectrum thus drops
by a factor of 28 between the Li and $\pi^0$ thresholds,
and continues to drop above the $\pi^0$ threshold,
offsetting the rise in the pion cross section.
This behavior thus suppresses $\pi^0$ production relative
to the GCR case, and thus we have a significantly higher
$\sigma_{\li6}^{\alpha \alpha}/\sigma_{\pi}^{pp}$ ratio.
As we will see, these ratios--and the differences between them--will 
be critical
in deriving quantitative constraints.

\section{Observational Inputs}
\label{sect:obs}

We have seen that the EGRB intensity and lithium abundances
are closely linked.  Here we collect
information on both  observables.

\subsection{The Observed Gamma-Ray Background and Limits
to the Pionic Contribution}
\label{sect:pionic}

Ever since $\gamma$-rays were first observed towards the Galactic
poles as well as in the plane \citep{fkh},
the existence of emission at high Galactic latitudes
has been regarded as an indication of an EGRB.
However, any information regarding the intensity, energy
spectrum, and even the existence of the EGRB
is only as reliable as the procedure for subtracting
the Galactic foreground.
Such procedures are unfortunately non-trivial and model-dependent.
The EGRET team \citep{sreekumar}
used an empirical model for tracers of Galactic hydrogen
and starlight, and found evidence for an EGRB which dominates
polar emission.
Other groups have recently presented new analyses of
the EGRET data.
In a semi-empirical approach using a model of Galactic
$\gamma$-ray sources, \citet{smr}
also find evidence for an EGRB,
but with a different energy spectrum
and a generally lower intensity than the
\citet{sreekumar} result.
Finally, \citet{kwl}
find that the Galactic foreground is sufficiently uncertain
that its contribution to the polar emission
can be significant, possibly saturating the observations.
Consequently, the \citet{kwl} analysis
is unable to confirm the existence of an EGRB in the EGRET
data; instead, they can only to place upper limits
on the EGRB intensity.

It was recently shown by \citet{prf} that a model-independent
limit on the fraction of EGRB flux that is of pionic origin (gamma rays that 
originate from $\pi^0$ decay) can be placed. Their limit comes from noticing 
that the EGRB shows 
no strong evidence of the  distinctive shape
pionic $\gamma$-ray spectral peak at $m_{\pi^0}/2$, the ``pion bump.''
Thus by comparing the shapes of the observed EGRB and 
theoretical pionic gamma-ray spectrum, they were able to maximize the pionic 
flux so that it stays below the observed one. This procedure allowed them to 
place constraints on the maximal fraction of EGRB that can be of pionic origin.

For the pionic $\gamma$-ray source-function, \citet{prf} used a 
semi-analytic fit from the \citet{ensslin} paper and the
\citet{dermer} model for the production cross section. 
A key feature of the pionic $\gamma$-ray spectrum is that
it approaches a power law at both high and low
energies, going to
$\epsilon^{\alpha_\gamma}$ for $\epsilon \ll m_{\pi}/2$
and to 
$\epsilon^{-\alpha_\gamma}$ for $\epsilon \gg m_{\pi}/2$.
In Dermer's model, the
$\gamma$-ray spectral index $\alpha_\gamma$
is equal to the cosmic-ray spectral index.
\citet{prf} adopted the value $\alpha_{\gamma}= 2.2$ for pionic 
extragalactic $\gamma$-rays, which is consistent with blazars and 
structure-forming cosmic rays as their origin. In this simple analysis, 
\citet{prf} used a single-redshift approximation, 
that is, they assumed that these $\gamma$-rays are all coming from one 
redshift, and thus their limit on the maximal pionic fraction is a function 
of $z$.

To obtain the EGRB spectrum from EGRET data, a  careful subtraction 
of Galactic foreground is needed. \citet{prf}  
considered two different EGRB spectra and obtained the following limit : 
for the  \citet{sreekumar} spectrum they found that 
the pionic fraction of the EGRB 
(integrated spectra above 100 MeV) can be as low as about 40\% for 
cosmic rays that 
originated at present, to about 
90 \% for $z=10$; for the more shallow spectrum of \citet{smr} they found 
that pionic fraction can go from about 30\% for $z=0$ up to about 70\% for 
$z=10$. However, the \citet{kwl} analysis of the EGRET data implies that 
the Galactic 
foreground dominates the $\gamma$-ray sky so that only an upper limit on the 
EGRB can be placed, namely $ I_{\gamma}(>100 \rm MeV) \le 0.5 \times 10^{-5} 
\ \ \rm cm^{-2} \ s^{-1} \ sr^{-1} $. 
Thus, in this case, we were not able to
obtain the pionic fraction.

However, to be able to connect the pionic $\gamma$-ray intensity 
$I_{\gamma}$ with lithium mole fraction $Y_i$ as shown in (\ref{eq:gamma2li}),
$I_{\gamma}$ must include all of the pionic $\gamma$-rays, that is, the 
spectrum has to be integrated from energy $\epsilon_0 = 0$. 
The upper limit to the pionic $\gamma$-ray
intensity above energy $\epsilon_0$ for a given redshift
can be written as
\begin{eqnarray}
I_{\gamma}(>\epsilon_0) 
  &=&  f_{\pi}(>\epsilon_0,z)I_{\gamma}^{\rm obs}(>\epsilon_0) \\
  &=&  \mathcal{N}_{\rm max} \int_{\epsilon_0} \varphi [\epsilon (1+z)] d\epsilon
\end{eqnarray} 
where $f_{\pi}(>\epsilon_0,z)$ is the upper limit to the fraction of 
pionic $\gamma$-rays \citep{prf}, $I_{\gamma}^{\rm obs}(>\epsilon_0) $ is the 
observed intensity above some energy, while $\varphi [\epsilon (1+z)] $ is 
the semi-analytic fit for pionic $\gamma$-ray spectrum \citep{ensslin} which 
is maximized with $ \mathcal{N}_{\rm max} $
normalization constant. An upper limit to the pionic $\gamma$-ray intensity
that covers all energies $I_{\gamma}(>0,z)$, follows immediately from the 
above equations:
\beq
\label{eq:I0}
I_{\gamma}(>0,z)=f_{\pi}(>\epsilon_0,z) I_{\gamma}^{\rm obs}(>\epsilon_0)
\frac{\int_0 \varphi [\epsilon (1+z)] d\epsilon}{\int_{\epsilon_0} \varphi 
[\epsilon (1+z)] d\epsilon}
\eeq

Now this is something that is semi-observational and  can be easily obtained 
from $\gamma$-ray intensity observed above some energy, and from \citet{prf} 
and \citet{ensslin} results.

\subsection{The Primordial Lithium Abundance}
\label{sect:primo}

Given the EGRB intensity, we will infer the amount of associated
lithium production.  
It will be of interest to compare this not only to the solar
abundance, but also to the primordial abundance of
\li7.
Metal-poor halo stars (extreme Population II)
serve as a ``fossil record'' of pre-Galactic lithium.
\citet{rbofn}
find
a pre-Galactic abundance
\beq
\label{eq:li-obs}
\pfrac{\rm Li}{\rm H}_{\rm pre-Gal,obs} = (1.23^{+0.34}_{-0.16}) \times 10^{-10}
\eeq
based on an analysis of very metal-poor halo stars.
On the other hand, one can use the WMAP \citep{wmap}
baryon density and BBN to predict a ``theoretical'' (or ``CMB-based'') 
primordial \li7 abundance \citep{cfo}:
\beq
\label{eq:li-thy}
\pfrac{\li7}{\rm H}_{\rm BBN,thy} = (3.82^{+0.73}_{-0.60}) \times 10^{-10}
\eeq

These abundances are clearly inconsistent.
Possible explanations for this discrepancy 
include
unknown or underestimated systematic errors in theory and/or observations
or new physics;
these are discussed thoroughly elsewhere 
\citep[see, e.g.,][and refs therein]{cfo4}.
For our purposes, we will acknowledge this
discrepancy by comparing pre-Galactic lithium production by
cosmic rays with both the observed and CMB-based Li abundances.

\section{\li6 and Gamma-Rays From Galactic Cosmic Rays}
\label{sect:gcr}

We have shown that \li6 abundances and 
extragalactic $\gamma$-rays are linked because
both sample cosmic-ray fluence,
and now apply this formalism to $\gamma$-ray and \li6 data.
In this section we turn to the hadronic products of 
Galactic cosmic rays, which are believed to be the dominant source of \li6,
but a sub-dominant contribution to the EGRB.

\subsection{Solar \li6 and Gamma-rays}

We place upper limits on the lithium component of GCR origin 
by using the formalism established in earlier sections.
To be able to find $I_{\gamma}/Y_{\rm ^6Li}$ from eq.~(\ref{eq:gamma2li})
we assume that ratio of cosmic-ray fluence along the line of sight 
(weighted by gas fraction)
to the local cosmic-ray fluence is $F_p(t)/F_p(\vec{x},t) \approx 1$.
That is,  we assume that the Milky Way fluence is
typical of star forming galaxies, i.e., that the $\gamma$-luminosities
are comparable:
$L_{\rm MW} \approx \langle L \rangle_{\rm gal}$.
Note that in the most simple case of a uniform approximation 
(cosmic-ray flux and gas fraction the same in all galaxies),
the two fluences would indeed be exactly equal.

Taking the solar $\li6$ abundance and 
$\langle \sigma_{\li6}^{\alpha \alpha} \rangle 
/ \langle \sigma_{\pi}^{pp} \rangle = 0.21 $ for  the ratio of GCR flux 
averaged cross-sections,
we can now use eq.~(\ref{eq:main}) to say that 
$I_{\gamma,\pi^0}(\epsilon >0)=9.06 \times 10^{-5} 
\rm cm^{-2} \ s^{-1} \ sr^{-1}$ is the hadronic $\gamma$-ray intensity that 
is required if all of the solar  $\rm ^6Li$ is made via Galactic cosmic-rays. 

We wish to compare this \li6-based pionic $\gamma$-ray flux
to the observed EGRB intensity 
$I_{\gamma}^{\rm obs}(\epsilon  >\epsilon_0)$. However, eq.~(\ref{eq:main})
gives the hadronic $\gamma$-ray intensity integrated over all
energies,
whereas the observed one 
is above some finite energy. Thus we have to compute
\beqar
\label{eq:galactic}
I_{\gamma,\pi^0}(\epsilon > \epsilon_0) 
   & = & I_{\gamma,\pi^0}(\epsilon >0) 
\frac{\int_{\epsilon_0} d \epsilon I_{\epsilon , \pi}}
     {\int_0 d \epsilon I_{\epsilon , \pi}} \\
   & = & 9.06 \times 10^{-5} {\rm cm^{-2} \ s^{-1} \ sr^{-1} }
\frac{\int_{\epsilon_0} d \epsilon I_{\epsilon , \pi}}
     {\int_0 d \epsilon I_{\epsilon , \pi}}
\eeqar
We follow the model of \citet{pf}
to calculate the GCR emissivity over the history of the universe.
The source function $q_\gamma^{\rm com}$ (equivalent to
eq.\ \ref{eq:qgamma}) is
given by a coarse-graining over galactic scales, so that
\beq
q_{\gamma,\rm gcr}^{\rm com}(z,\epsilon) 
  = L_\gamma(\epsilon) n_{\rm gal}^{\rm com}(z)
\eeq
where $L_\gamma$ is the average galactic $\gamma$-ray luminosity
(by photon number),
and $n_{\rm gal}^{\rm com}(z)$ is the mean comoving number density of
galaxies.  
The key assumptions for the luminosity 
$L_\gamma$ are: (1) that supernova explosions provide
the engines powering cosmic-ray acceleration,
so that the cosmic-ray flux $\Phi \propto \psi$
scales with the supernova rate and thus the star formation rate
$\psi$; (2) that the targets come from the gas mass which
evolves following the ``closed box'' prescription; and
(3) that the Milky Way luminosity represents that of an average
galaxy.
With these assumptions
we have that $L_\gamma \propto \mu \psi$, and
thus that $q_\gamma^{\rm com} \propto \mu \dot{\rho}_\star$,
where $\dot{\rho}_\star$ is the cosmic star formation rate.

Following \citet{pf}, the specific form of $I_{\epsilon , \pi}$
is expressed in terms of the present day Milky Way gas mass
fraction $\mu_{0,\rm MW}$, cosmic star-formation rate 
$\dot{\rho}_\star(z)$, Milky Way
gamma-ray (number) luminosity $L_{\gamma , \rm MW}(z,E)$, cosmology 
$\Omega_{\Lambda}$ and $\Omega_m$, and integrated up to
$z_*$, the assumed starting redshift for star formation. For this calculation
we adopt 
the following values: $\mu_{0,\rm MW}=0.14$, $\Omega_{\Lambda}=0.7$,
 $\Omega_m=0.3$ and $z_*=5$. For the 
cosmic star formation rate we use the dust-corrected analytic fit 
from 
\citet{cole}. Finally we need the (number) luminosity of pionic gamma-rays 
which we can write as
\beq
L_{\gamma ,\rm MW}(z,E)= \Gamma_\gamma N_p = \frac{q_{\gamma, \pi}}{n_p}N_p
 \propto \Phi M_{\rm gas}
\eeq
where $n_p$ is the proton number density in the Galaxy, $N_p$ is the total
number of protons in the Galaxy, while $q_{\gamma, \pi}[\rm s^{-1} GeV^{-1}
cm^{-3} sr^{-1}] $ is the
source function of gamma-rays that originate from pion decay adopted from
\citet{ensslin}. Notice that in equation (\ref{eq:galactic}) we have
the ratio of two integrals where integrands are identical, thus normalizations
 and constants will cancel out.
Therefore, instead of using the
complete form of $L_{\gamma , \rm MW}(z,E) $ we need only use the spectral shape
of the pionic gamma-ray source function \citep{ensslin}, that is, only the 
part that is energy-and redshift-dependent. 

Finally then, we find 
\beq
\label{eq:gamlim}
I_{\gamma,\pi^0}(\epsilon > 0.1 \rm GeV)
  = 3.22 \times 10^{-5} \rm \ cm^{-2} \ s^{-1} \ sr^{-1}
\eeq
which we can now compare to the observed EGRB values  
$I_{\gamma}^{\rm obs}(\epsilon  >0.1 \rm GeV)$ that are given in the first column
of Table \ref{table:res}. As one can see, our {\em pionic} 
EGRB gamma-ray intensity
is between 2 and 6 times larger than the {\em entire} observed value!

We thus conclude that the solar \li6 abundance, if made
by GCRs as usually assumed, seems to demand
an enormous diffuse pionic $\gamma$-ray contribution,
far above the entire EGRB level.
How might this discrepancy be resolved?
One explanation follows by dropping our assumption that
$F(\vec{x}_{\rm MW},t_0) = F_{\rm avg}(t_0)$, i.e., that
the baryon-weighted Milky Way GCR fluence is the same as the
cosmic mean for star-forming galaxies.
Note that we have $F = \int dt \mu \Phi 
 \propto  \int dt \, \langle \psi M_{\rm gas} \rangle$,
where $\psi$ is the global galactic star formation rate
(assuming $\Phi \propto \psi$),
and $M_{\rm gas}$ the galactic gas content.
If our Galaxy has an above-average star formation rate
and/or gas mass, this will increase the local
\li6 production relative to the average over all
galactic populations,
and thus lead to an overestimate  of
the EGRB.

In this connection it is noteworthy to compare
our \li6-based estimate of the galactic EGRB contribution
to the work of \citet{pf}.  That calculation adopted
the same model for the redshift history of cosmic-ray
flux and interstellar gas, and so only differed from
the present calculation in the normalization
to Galactic values.
\citet{pf} normalized to the {\em present} Galactic $\gamma$-ray
luminosity.  This amounts to a calibration not to
the time-integrated cosmic-ray {\em fluence}, but rather to the
instantaneous cosmic ray {\em flux}, 
as determined by the 
\citet{dermer} emissivity, a Galactic gas
mass of $10^{10} M_\odot$, and an estimate of the
present Galactic star formation rate.
This normalization gave a galactic EGRB component
which at all energies lies {\em below} the
total \citep{sreekumar} background.
Our calculation is normalized to solar \li6,
which is a direct measure of Galactic (or at least solar neighborhood)
cosmic-ray {\em fluence},
and which contains fewer uncertainties
than the factors entering in the \citet{pf}
result.
Yet surprisingly, the \li6-based fluence result
gives a high pionic EGRB, while the 
more uncertain normalization gives an acceptable result.

Can we independently test whether our Galaxy has
had an above-average cosmic-ray exposure?
This present a challenge,
as we require an integral measure of cosmic-ray
activity, which is readily available locally but
difficult to obtain in external galaxies.
The best candidates are the LiBeB isotopes;
\li6 is ideal for the reasons we have outlined, but 
is not accessible in stars bright enough to be see
in even the nearest external galaxies.
The best hope then would be for measurements of
\iso{Be}{9}, in the Local Group or beyond.
In the SMC, such measurements have placed interesting
limits on boron abundances
\cite{brooks},
though the presence of the neutrino-process
production of \bor11 
makes boron observations more difficult to interpret
than beryllium.

Another explanation 
for the high intensity of eq.~(\ref{eq:gamlim}) 
stems from noting that 
the required ERGB intensity is much larger for GCRs than that
one would infer from SFCRs, due to the
large difference in the $\pi^0/\li6$ ratio for these
two spectra.  
Were the Milky Way spectra is atypically skewed to high energies,
we would overestimate the $\pi^0/\li6$ ratio 
and thus the EGRB contribution.
This possibility (which we regard as less likely than
the previous one) could be tested
by observations of cosmic-ray spectra in 
external Galaxies, e.g., by $\gamma$-ray observations of
Local Group galaxies such as GLAST should perform
\cite{pf1}.

Finally, a related but more unconventional  view would be 
that \li6 is in fact primarily made  by SFCRs themselves,
rather than by GCRs.
This suggestion is further discussed and constrained
below, \S \ref{sect:sfcr}.

We close this subsection by noting that if the depletion of \li6
is taken into consideration, one might use \li6  abundance  larger
than solar. In that case one would find that the accompanying {\em pionic} 
EGRB gamma-ray intensity is more  than 2-6 times greater than the observed 
EGRB.

\subsection{The Observed EGRB and Non-Primordial \li6 }

We can exploit equation (\ref{eq:main}) in both directions.
Here we use the observed EGRB spectrum to constrain the \li6 
abundance produced via Galactic cosmic rays.
By comparing this Galactic \li6 component to the observed solar abundance 
we can then place an upper limit on the {\em residual }
\li6 which (presumably) was produced by SFCR. 
As described in \S\ref{sect:pionic}, with the
observed EGRB spectrum in hand we can
place an upper limit on its fraction of pionic origin. In the
case of SFCR-produced pionic gamma-rays, we can place
constraints directly only in the presence of a model for
the SFCR redshift history.
Since a full model is unavailable,
below (\S \ref{sect:sfcr}) we adopt the ``single-redshift 
approximation.''
However, in the case of galactic cosmic rays we have a better
 understanding of 
the redshift history of the sources.
Therefore, we will follow \citet{pf} to calculate the pionic differential 
gamma-ray intensity for some set of energies
\begin{eqnarray}
\nonumber
I_{\gamma_{\pi,E}} &=&
\frac{c}{4\pi H_0 \psi_{\rm MW}} \int_0^{z_*} dz 
\frac{ \dot{\rho}_\star(z) L_{\gamma_{\pi}} [(1+z)E]}{\sqrt{\Omega_{\Lambda}
+\Omega_M (1+z)^3 }}  \\ 
 &  & \ \ \ \ \times \left[ \frac{1}{\mu_{0,\rm MW}}-\left( 
\frac{1}{\mu_{0,\rm MW}} -1 \right) \frac{\int_{z_*}^z dz (dt/dz)\dot{\rho}_\star(z)}{
\int_{z_*}^0 dz (dt/dz)\dot{\rho}_\star(z)} \right] 
\label{eq:gcr_pi_flux}
\end{eqnarray}
where $I_{\gamma_{\pi}}$ is in units of $\rm s^{-1} \ cm^{-2} \ GeV^{-1}$ ,
and $\psi_{\rm MW}$ is the present Milky Way star formation rate.
For the pionic gamma-ray luminosity
$L_{\gamma_{\pi}}$ we will, as before, use the pionic gamma-ray source function
adopted from \citet{ensslin} ($\alpha_{\gamma}=2.75$ for GCR spectrum), 
however we will let the normalization be determined by
maximizing the pionic contribution to the EGRB. 
The adopted parameters, cosmology, and cosmic star formation rate we
keep the same as in previous subsection.

\begin{figure}[t]
\epsscale{0.5}
\plotone{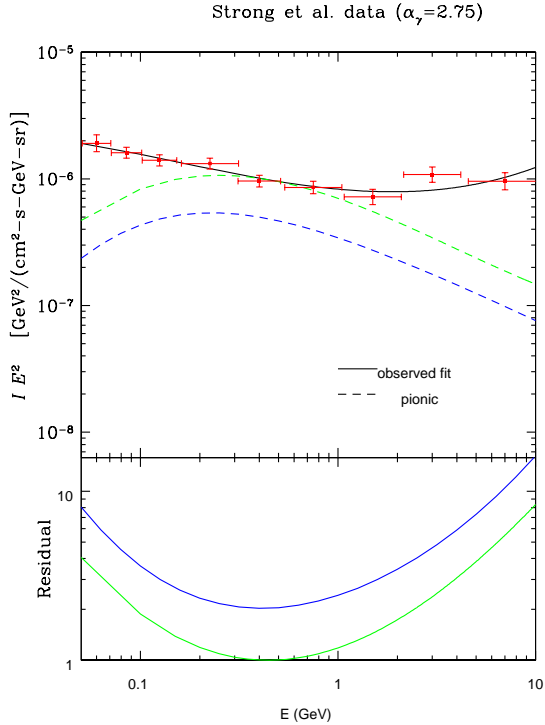}
\caption{In the upper panel of this figure, we plot the pionic 
(dashed lines:green- maximized, blue- normalized to the Milky Way)  
EGRB spectrum, where decaying pions are of GCR origin, compared to the observed
EGRB spectrum (solid line, fit to data); for purposes of illustration, 
we use the
\citet{smr} data points, which are given in red crosses.
The bottom panel represents the residual
function, that is, 
$\log [(I E^2)_{\rm obs}/(I E^2)_{\pi}]=\log (I_{\rm obs}/I_{\pi})$.}
\label{fig:fraction}
\end{figure}

Once we obtain the spectrum we can then fit it with
\beq
\ln ( I_{\gamma_{\pi}}E^2)=-14.171-0.546 \ln E -0.131 (\ln E)^2+0.032 (\ln E)^3
\eeq
where $E$ is in GeV and $I$ is in
$\gamflux \ {\rm GeV^{-1}}$.
The free leading term in the above equation is set by requiring that 
$ I_{\gamma_{\pi}}=I_{\gamma,\rm obs}$ at the energy $E=0.44$ GeV which maximizes
pionic contribution by demanding that the pionic gamma-ray spectrum always 
stays below the observed one (since the feature of pionic peak is 
not observed).
We also fit the \citet{smr} data with
\beq
\ln ( I_{\gamma,\rm obs} E^2)=-14.003-0.144 \ln E -0.097 (\ln E)^2+0.017 (\ln E)^3
\eeq
in the same units.

By going through the procedure described in \S\ref{sect:pionic} 
\citep[see][for more detail]{prf} 
we can now obtain an upper limit to the fraction of pionic
gamma-ray compared to the \citet{smr} observed EGRB spectrum. This maximized 
pionic (green dashed line), as well as the observed, gamma-ray spectrum is 
presented in 
Fig.~\ref{fig:fraction}.
We find the upper limit to pionic fraction to be
$f_{\pi}(>0.1 {\rm GeV}) \equiv \int_{0.1}dE \, I_{\gamma_{\pi}}/ 
\int_{0.1}dE \, I_{\gamma,\rm obs}=0.75 $. 
We note in passing that a maximal pionic fraction
as appears in 
Fig.~\ref{fig:fraction}
gives a poor fit at energies both above and below
the matching near 0.4 GeV, suggesting the presence of
other source mechanisms. This mismatch reflects a
similar problem in the underlying Galactic
$\gamma$-ray spectrum, and suggests that the pionic contribution
to the EGRB is in fact sub-maximal.

Thus, the pionic gamma-ray flux above
0.1 GeV is $I_{\gamma_{\pi}}(>0.1 \ {\rm GeV})=0.83 \times 10^{-5} \rm 
cm^{-2} \ s^{-1} \ sr^{-1} $.
From eq.~(\ref{eq:galactic}) it now follows that the total flux 
is
 $I_{\gamma_{\pi}}(>0)=2.31 \times 10^{-5} \rm cm^{-2} \ s^{-1} \ sr^{-1} $.
As before, we can now use eq.~(\ref{eq:main}) to find the GCR \li6 mole 
fraction
\beq
\left( \frac{Y_{\li6}}{Y_{\li6_{\odot}}} \right)_{\rm GCR}=
\frac{I_{\gamma_{\pi}}(>0)}{9.06 \times 10^{-5}\ \gamflux}=0.25
\eeq
and thus, SFCR-produced \li6 can be at most (neglecting the \li6
depletion) the residual \li6
\beqar
 \left( \frac{Y_{\li6}}{Y_{\li6_{\odot}}} \right)_{\rm SFCR} &=&
1-\left( \frac{Y_{\li6}}{Y_{\li6_{\odot}}} \right)_{\rm GCR}  \\ 
 &=& 0.75
\nonumber
\eeqar
With the appropriate scaling between \li7 and \li6 as given in Table
\ref{table:scaling}, we can then determine the total elemental
${\rm Li}  = \li7+\li6 $
abundance and compare it to the primordial values from (\ref{eq:li-obs}) and
(\ref{eq:li-thy}): 
\beq
\nonumber
\pfrac{\rm Li}{\rm H}_{\rm SFCR}
 = 3.45 \times 10^{-10}
 = 0.90 \pfrac{\li7}{\rm H}_{\rm p,thy}
 = 2.81 \pfrac{\li7}{\rm H}_{\rm p,obs}
\eeq

So far we have been determining the maximized pionic fraction of the EGRB 
based only on the shape of the pionic spectrum. However, in the case of 
normal galaxies we have a better understanding of what that fraction should be.
That is, we can normalize pionic spectrum to the one of the Milky Way, and
then integrate over the redshift history of sources. Following \citet{pf} (and
references therein) we set up the normalization by requiring that
$\int_{0.1 \ \rm GeV} dE L_{\gamma_{\pi}}(z=0,E)= 
\int_{0.1 \ \rm GeV} dE L_{\gamma_{\pi},\rm MW}(E)=
2.85 \times 10^{42} \ \rm s^{-1}$. With the analytic fit of the shape of the 
pionic spectrum from \cite{ensslin} this now gives:
\beq
L_{\gamma_{\pi}}(z=0,E)=9.52 \times 10^{44} \rm \ s^{-1} \ GeV^{-1}
\left[ \left( \frac{2\epsilon}{m_{\pi^0}} \right)^{\delta_{\gamma}} 
 + \left( \frac{2\epsilon}{m_{\pi^0}} \right)^{-\delta_{\gamma}} \right]^{-\alpha_{\gamma}/\delta_{\gamma}}
\eeq
where $\alpha_{\gamma}=2.75$ for GCR spectrum, and $\delta_{\gamma}=
0.14 \alpha_{\gamma} ^{-1.6}+0.44$. Now we can use equation 
(\ref{eq:gcr_pi_flux}) to obtain the pionic spectrum which  is plotted on the 
Fig.~\ref{fig:fraction}(blue dashed line).We use star formation rate
$\psi_{\rm MW}=3.2 M_{\odot} \ \rm yr^{-1}$ \citep{mckee}.
Finally we find that in this case, when pionic spectrum is normalized to the
Milky Way, the GCR $\li6$ mole fraction that accompanies it is
\beq
\nonumber
\left( \frac{Y_{\li6}}{Y_{\li6_{\odot}}} \right)_{\rm GCR}=0.14
\eeq
which then gives
$({\rm Li/H})_{\rm SFCR}=3.96 \times 10^{-10}
 =1.03 ({\li7}/{\rm H})_{\rm p,thy}=
3.22 (\li7/{\rm H})_{\rm p,obs}$,
which is of course a weaker limit than the 
maximal pionic case.

We thus see that 
in a completely model-independent analysis,
current observations allow the 
possibility that SFCRs are quite a significant source of 
\li6 and of $\gamma$-rays.
Indeed, we cannot exclude that SFCR-produced
lithium can be a potentially large contaminant
of the pre-Galactic Li component of halo stars,
which would exacerbate the already troublesome
disagreement with CMB-based estimates of primordial \li7.
Consequently, we conclude that
models for SFCR acceleration and propagation should
include both $\gamma$-ray and \li6 production; and
more constraints on SFCR, both theoretically (e.g., 
space and time histories) 
and observationally (e.g., EGRB and possibly diffuse synchrotron
measurement), will clarify the picture we have sketched.

Note that had we also considered the possibility of
depletion of \li6 , we would have found a 
greater 
\li6 residual, and thus had an even larger SFCR-produced component.

\section{\li6 and Gamma-Rays From Cosmological Cosmic Rays}
\label{sect:sfcr}

In this section we turn to the as-yet unobserved
cosmological component of cosmic rays, and  to the synthesis of lithium by 
SFCR.
This lithium component would be the first
made after big bang nucleosynthesis.
Any Li which is produced this way prior to
the most metal-poor halo stars
would amount to a pre-Galactic Li enrichment
and thus would be a {\em non-primordial} Li component,
unaccompanied by beryllium and boron production.
This structure-formation Li would be an additional ``contaminant''
to the usual components in halo stars, the 
\li7 abundance due to primordial nucleosynthesis, and
the \li6 and \li7 contribution due to Galactic cosmic rays
\citet{rbofn}.
Moreover, the pre-Galactic but non-primordial component
would by itself be indistinguishable from the true primordial
component, and thus would lead to an overestimate of
the BBN \li7 production.

Our goal in this section is to exploit the $\gamma$-ray
connection to constrain the structure-formation Li contamination.
Unfortunately, we currently lack a detailed  understanding of the amount and 
time-history of the structure formation cosmic rays (and resulting
$\gamma$-rays and Li).  Thus we will make the conservative assumption
that {\em all} structure formation cosmic rays, and the resulting
$\gamma$s and Li, are generated prior to {\em any} halo stars.
Furthermore, we will assume that the pionic contribution
to the EGRB is {\em entirely} due to structure formation cosmic rays.
This allows us to relate observational limits on the pionic EGRB
to pre-Galactic Li.

With this assumption and a SFCR composition
$\Phi_{\alpha}^{\rm cr}/\Phi_p^{\rm cr} \approx y_{\alpha}^{\rm cr} = 0.1$,
we can now use the
appropriate scaling factor from Table \ref{table:scaling} to rewrite 
eq.~(\ref{eq:gamma2li}) 
\beqar
\label{eq:sfcr_convert}
I_{\gamma,\pi^0}(\epsilon >0, z)
 & = & \frac{\xi_\gamma \zeta_\alpha}{4\pi y_{\rm \alpha}^{\rm cr} 
y_{\rm \alpha}^{\rm ism}} 
  \frac{\zeta \sigma_{\pi^0}}{\sigma_{\li6}^{\alpha \alpha}} 
\pfrac{\li6}{\rm H} n_{\rm b} c \\
  & = & 1.86 \times 10^{-5} \ \gamflux \ 
      \pfrac{\li6}{\li6_\odot} 
\eeqar
or
\beq
\label{eq:li6}
\pfrac{\li6}{\li6_\odot} = 0.538 \pfrac{I_{\gamma,\pi^0}(>0)}
{10^{-5} \ \rm cm^{-2} \ s^{-1} \ sr^{-1}}
\eeq
where we used the solar lithium mole fraction  
$Y(\li6)_{\odot}=1.09 \times 10^{-10}$.

To set up an extreme upper limit on pre-Galactic SFCR \li6,
we assume that the {\em entire} pionic extragalactic gamma-ray background 
came from SFCR-made pions,
and was created prior to {\em any} halo star.  
As mentioned in the previous section, the method used in subtraction of the 
Galactic foreground is crucial for obtaining the EGRB spectrum. What is more, 
the EGRB spectrum is an important input parameter in 
the \citet{prf} analysis
whose estimates of the maximal pionic gamma-ray flux we will use here.
Our results for the SFCR lithium upper limits are collected in 
Table \ref{table:res}. The results depend on the choice of the EGRB 
spectrum as well as the redshift of origin of cosmic-rays according to the 
single-redshift approximation used by \citet{prf} to obtain the maximal 
pionic EGRB fraction.  Note that we considered only the two most extreme 
redshifts to illustrate the results.
In the Table \ref{table:res}, $z$ is the redshift, $I_{\gamma,\pi}(>0)$ is the
upper limit for the pionic $\gamma$-ray intensity above 0 energy determined 
from
(\ref{eq:I0}) as explained in the previous section, 
$({\rm Li/H})_{\rm SFCR}^{\rm max}$
is the upper limit to total ($\li6+\li7$) lithium abundance that can 
be of SFCR origin, while ${\rm Li}_p^{\rm theo}$ and  
${\rm Li}_p^{\rm obs}$ 
are the theoretical and observational
primordial lithium abundances respectively as given in equations
(\ref{eq:li-thy}) and (\ref{eq:li-obs}).

Notice that for the case of \citet{kwl} EGRB, since a spectrum was unavailable, 
the procedure described in the section \S \ref{sect:pionic} for maximizing the pionic fraction 
of the EGRB could not be used. Thus, to place an upper limit on SFCR lithium we
assumed that the  entire EGRB can be attributed to decays of pions, that is, assume $I_{\gamma}=I_{\gamma,\pi^0}$.
For the \citet{sreekumar} and \citet{smr} EGRB spectra, we use 
the upper limits
to $I_{\gamma,\pi^0}$ obtained by \citet{prf}.
Once the $I_{\gamma,\pi^0}$ is set we can use (\ref{eq:li6}) to find the
SFCR $\li6$ upper limit.

To find the total halo star contribution we must also include
\li7, which is in fact produced more than
\li6 in $\alpha \alpha$ fusion:
as seen in Table \ref{table:scaling},
${\rm (^7Li/\li6)}_{\rm SFCR} = \langle \sigma_{^7Li}^{\alpha \alpha} \rangle 
/ \langle \sigma_{\li6}^{\alpha \alpha} \rangle \approx 2$.
The total SFCR elemental Li production 
appears in Table \ref{table:res}, both in terms of
the absolute Li/H abundance and its ratio to 
the different measures of primordial Li (\S\ref{sect:primo}).

\begin{table}[h]
\epsscale{0.5}
\begin{center}
\caption {Upper limit on Li of SFCR origin \label{table:res}}
\begin{tabular}{ccccccc}
\tableline\tableline
EGRB  [$\rm cm^{-2} s^{-1} sr^{-1}$]& $z $ & $I_{\gamma ,\pi}(>0)$ & 
$\rm (Li/H)_{SFCR}^{max} $ & 
$ \rm \frac{Li_{SFCR}^{max}}{Li_p^{theo}} $ & 
$ \rm  \frac{Li_{SFCR}^{max}}{Li_p^{obs}} $ & $f_\pi $ \\
\tableline\tableline
\citet{sreekumar} & 0 & $8.78\times 10^{-6} $ & $ 2.19 \times 10^{-10} $ 
& 0.57 & 1.78 & 0.91	\\
 $I_{\gamma , \rm obs} (>0.1)= 1.57 \times 10^{-5} $
  & 10 & $1.22\times 10^{-4} $ & $ 3.04 \times 10^{-9} $ & 7.95 & 24.7 
& 0.15	 \\
\tableline
Strong et al. (2004) & 0 & $4.59\times 10^{-6} $ & $ 1.14 \times 10^{-10} $ & 
0.30 & 0.93 & 1.29	 \\
$I_{\gamma , \rm obs} (>0.1)= 1.11 \times 10^{-5} $ 
& 10 & $6.27\times 10^{-5} $ & $ 1.56 \times 10^{-9} $ & 4.09 & 12.69 
& 0.21	\\
\tableline
Keshet et al. (2003)  & 0 & $ < 6.5\times 10^{-6} $ & $ < 1.62 \times 10^{-10} $ 
& 0.42 & 1.32 & 2.86 	\\
 $I_{\gamma , \rm obs} (>0.1) < 0.5 \times 10^{-5} $ 
& 10 & $ < 4.03\times 10^{-5} $ & $ < 1.00 \times 10^{-9}$ & 2.63 & 8.16  
& 0.46	\\ 
\tableline
\end{tabular}
\end{center}
\end{table}

From Table \ref{table:res} we see that the 
maximal possible SFCR contribution to halo star lithium
could be quite substantial.  
If the pre-Galactic SFCR component is dominantly produced at 
high redshift (i.e., as in the $z \sim 10$ results) then
the maximum allowed Li production is can exceed the
primordial Li production (however it is estimated),
in some cases by a factor up to 25!  
The situation is somewhat better if the pre-Galactic SFCR
production is at low redshift, but here it is hard to
understand how this would predate the halo star component of
our Galaxy.
The high-redshift result is thus the more likely one,
but also somewhat troubling in that the limit
is not constraining.  
The indirect limits on SFCR Li in the previous section are
somewhat stronger, but these also hold the door open
for a significant level of pre-Galactic synthesis.

We caution that the lack of a strong constraint on 
SFCR Li production is not the same as positive
evidence that the production was large.
Recall that we have made several assumptions
which purposely maximize the SFCR contribution;
to the extent that these assumptions fail, the
contribution falls, perhaps drastically.
A more detailed theoretical and observational understanding
of the SFCR history, and of the 
EGRB, will help to clarify this situation.
Moreover, given that the
halo star Li is already found to be below the
CMB-based \li7 BBN results, we are already strongly
biased to believe that the pre-Galactic SFCR component
is {\em not} very large.
Thus one might be tempted instead to go the other way
and use Li abundances to constrain SFCR activity.

We thus now go the other way and use solar \li6 to constrain the SFCR 
$\gamma$-ray flux. Again, given our incomplete knowledge of SFCRs,
we must adopt a simplifying assumption
about the degree of \li6 production which is due to SFCR.
To be conservative, we make 
the extreme assumption is that {\em all} of the solar \li6 is produced 
by SFCR, and thus find via eq.\ \pref{eq:sfcr_convert} 
that $\gamma$-flux is $I_{\gamma,\pi^0} (>0 \rm \ GeV) > 1.86 \times 10^{-5} \ \gamflux $. From (\ref{eq:I0}) we can determine $I_{\gamma,\pi^0} (>0.1 \rm \ GeV)$ to be $0.23 \times 10^{-5} <I_{\gamma,\pi^0} (>0.1 \rm \ GeV)<1.43 \times
10^{-5}\gamflux $ depending on the redshift of pionic $\gamma$-rays,
 which is below the observed level
as determined by \citet{sreekumar}, and
a factor of 2-14 lower than the prediction based
on GCR. Thus, for a given observed intensity
$I_{\gamma}^{\rm obs}(>\epsilon_0)$ we can now use (\ref{eq:I0}) to constraint
the hadronic fraction of EGRB, that is, calculate $f_{\pi}(>\epsilon_0 , z)$ which
is also presented in the  Table \ref{table:res}.
 
However, since Li is being
depleted, the use of the solar \li6 abundance does not give us the upper most
limit to the required pionic gamma-ray flux $I_{\gamma,\pi^0}(>0)$. Thus,
if one would to compensate for the depletion, the pionic fraction 
$f_{\pi}(>\epsilon_0 , z)$ would become even larger.

Indeed, this may suggest a solution to the 
EGRB overproduction by GCRs, seen in the previous section.
If \li6 is mostly made by SFCRs, then
the associated $\gamma$-ray production is in line
with the observed background.
In this case, \li6 would still be of cosmic-ray origin, but
not dominated by GCR production.
Such a scenario faces tests regarding 
\li6 and other LiBeB abundances and their
Galactic evolution.
A detailed discussion of this scenario 
will appear in a forthcoming study.

\section{Discussion}
\label{sect:discuss}

The main result of this paper is to identify and
quantify the tight connection between \li6
and the EGRB as measures of cosmic-ray
history.
Specifically, these two observables provide measures
of average, gas-weighted cosmic-ray fluence.
Moreover, the observables are
complementary, in that \li6 samples local
fluence, while the EGRB encodes the cosmic mean fluence.

We present scaling laws which relate \li6 and
the EGRB intensity for different cosmic ray spectra
appropriate for GCR and SFCR populations.
Using these scalings, and assuming that
our local \li6 measurements are typical,
we can test the self-consistency of \li6 and EGRB observations 
in a relatively model-independent manner.  
We find that if SFCR dominate the pionic
EGRB, then the associated \li6 production can
be a significant and perhaps dominant contribution
to the solar abundance.
On the other hand, we find that if \li6 production is
dominated by GCRs, then the associated
$\gamma$-ray production is enormous, at least
a factor of two above the observed intensity.

Furthermore, using the EGRB
we use two different lines of argument to 
place an upper limit on the SFCR contribution
to pre-Galactic lithium in halo stars.
Such a component of lithium would be confused
with the true primordial abundance and thus
would exacerbate the existing deficit in
halo star Li relative to the CMB-based
expectations of BBN theory.
Unfortunately, current EGRB data are such that
our model-independent upper limit (which most assume,
among other things, that {\em all} SFCRs are created
prior to {\em any} halo stars)
is very weak.  In particular, we cannot exclude
the possibility that a significant portion of pre-Galactic 
lithium is due to SFCRs.
We thus find that the nucleosynthesis aspects
of SFCRs are important and deserver further
more detailed study.

A full understanding of the 
implications of the relationships among 
\li6, diffuse $\gamma$-rays, and cosmic
ray populations thus awaits
better observational constraints (both light elements
and especially the EGRB) as well as a more detailed
study of SFCRs.
Having shown the importance of both the \li6 and $\gamma$-ray constraints,
it is our hope that these observables will be 
calculated in models of galactic and structure formation
cosmic rays, and that both \li6 and the EGRB
are used in concert to constrain 
cosmic-ray interactions with diffuse matter.
The results of these models will
go far to address some of the questions
which this study has raised.

\acknowledgments
We are grateful for illuminating discussions
with Vasiliki Pavlidou.
This material is based upon work supported by the National Science
Foundation under Grant No.~AST-0092939.

\appendix

\section{Notation and Normalization Conventions}
\label{sect:norm}

The interactions of cosmic-ray species $i$ with
target nucleus $j$ produces species $k$ 
at a rate per target particle of
\beq
\label{eq:rate}
\Gamma_k = \int_{E_{{\rm th},k}} dE \, \sigma_{ij \rightarrow k}(E) \phi_i
  \equiv \sigma_{ij \rightarrow k} \Phi_i
\eeq
Here $E$ is the cosmic-ray energy per nucleon,
$\sigma_{ij \rightarrow k}$ is the energy-dependent production
cross section, with threshold $E_{{\rm th},k}$, and
$\phi_i$ is the cosmic-ray flux.
The rate per unit volume for $i+j \rightarrow k$ is thus
$q_k = \Gamma_k n_j$.

Note that the flux in eq.~(\ref{eq:rate})
is position- and time-dependent.
To isolate this dependence, it is useful to
define a total, energy-integrated, flux
\beq
\label{eq:totflux}
\Phi_i = \int_{E_{\rm th,min}} dE \, \phi_i
\eeq
where we choose the lower integration limit to
always be the {\em minimum} threshold $E_{\rm th,min}$ for
all reactions considered; in our case this
is the $\alpha + \alpha \rightarrow \li7$ threshold
of $8.7$ MeV/nucleon.
From eqs.~(\ref{eq:rate}) and (\ref{eq:totflux})
it follows that 
\beq
\sigma_{ij \rightarrow k} = \Gamma_k/\Phi_i
\eeq
represents a flux-averaged cross section.
Also note that if the spectral {\em shape} of
$\phi_i$ is constant (as we always assume), then
so is $\sigma_{ij \rightarrow k}$,
and the flux $\Phi_i$ contains all of the time and space 
variation of $\Gamma_k$.

Finally, two conventions are useful for quantifying abundances.
Species $i$, with number density $n_i$,
has a ``mole fraction'' (or baryon fraction) 
$Y_i = n_i/n_{\rm b}$.
It is also convenient to introduce the
``hydrogen ratio'' $y_i = n_i/n_{\rm H} = Y_i/Y_{\rm H}$.

\section{Cosmic Gamma-Radiation Transfer}
\label{sect:radtransf}

The expression for $\gamma$-ray intensity in a
Friedmann universe is well-known \citep{stecker}, 
but usually expressed in redshift space.
For our purposes, the result expressed in the time domain
is critical, and indeed is more fundamental,
so we give a derivation based on the Boltzmann
equation.
For this section we adopt units in which $c=1$.

The differential photon (number) intensity $I$
is directly related via
\beq
\label{eq:f2I}
I(\vec{p},\vec{x},t) = p^2 f(\vec{p},\vec{x},t)
\eeq
to the $\gamma$-ray distribution function
$f(\vec{p},\vec{x},t) = d^3N/d^3 \vec{p} d^3\vec{x}$
Here $\vec{p}$ and $\vec{x}$, as well as the
volume elements, are physical quantities
(and thus subject to change with cosmic expansion).
The distribution function is related to the 
photon sources via the relativistic Boltzmann equation
\beq
p^\mu \pd_\mu f - \Gamma_\mu^{\alpha \beta} p^\alpha p^\beta f
  = E \frac{dq}{d^3 \vec{p}} 
\eeq
where gravitational effects enter through the
Affine connection $\Gamma$, 
where $E = p = |\vec{p}|$,
and where the source function (number of photons created per unit
volume per unit time) is $q$.
 
For an isotropic FRW universe we have
$f = f(E,t)$, and thus
\beq
\pd_t f - \frac{\dot{a}}{a} E \pd_E  f = \frac{q(E)}{4\pi E^2}
\eeq
where $q(E) = dq/dE$
and where we neglect photon scattering and absorption.

We now note that a given photon's energy
$E$ drops due to redshifting as $a^{-1}$.
It is thus useful to define a comoving energy
$\epsilon = a E$;
with $a(t_0) = 1$, we see 
that $\epsilon$ is also the {\em present-day (observed) photon energy}.  
Changing variables
from $f(E,t)$ to $f(\epsilon,t)$, and similarly for $q$, 
the energy-dependent $\partial_\epsilon$ term drops out;
this is physically reasonable since we do not allow for
scattering processes, and thus a photon's 
energy can only change due to redshifting.
We 
then have
\beq
\pd_t f = a^2 \frac{q(\epsilon/a)}{4\pi \epsilon^2}
\eeq
which, for any fixed comoving energy $\epsilon$, integrates to
\beq
f(\epsilon,t) = \frac{1}{4\pi \epsilon^2} 
  \int_0^t dt^\prime \ a^2 \ q(\epsilon/a)
\eeq

Equation \pref{eq:f2I} then gives the intensity
\beq
\label{eq:spectrum}
I(\epsilon,t)  = \frac{1}{4\pi a(t)^2} 
  \int_0^t dt^\prime \ \frac{1}{a(t^\prime)} q_{\rm com}[\epsilon/a(t^\prime)]
\eeq
where $q_{\rm com} = a^3 q$ is the comoving
source rate.  Equation \pref{eq:spectrum}
is the usual expression (which is often
then expressed in terms of an integral over
redshift).
Finally, if we integrate over the entire energy
spectrum, and evaluate at the present epoch $t_0$ (when $a_0 = 1$),
we have
\beq
\label{eq:totalgamma}
I(>0,t_0) = \int_0^\infty d\epsilon \, I(\epsilon,t_0)
 =  \frac{1}{4\pi} \int_0^{t_0} dt^\prime \ q_{\rm com}(>0)
\eeq
where $q_{\rm com}(>0) = \int_0^\infty du \ q(u)$ is
the total source rate, integrated over rest-frame energy.

We see from
eq.\ \pref{eq:totalgamma} 
that the energy-integrated intensity 
is the same as one would find from uniform sources
in a non-expanding universe (which have been ``switched on'' for
a duration $t$).  
This result is physically sensible, because the two
effects of cosmic expansion are to 
introduce a particle horizon and redshifting.
The energy integration removes the effect of
redshifting, so that the only effect is that of the particle
horizon, which acts to set the integration timescale.

\end{document}